\documentclass[sigconf,screen]{acmart}

\setcopyright{acmcopyright}
\acmYear{2021}
\copyrightyear{2021}
\acmConference[ESEC/FSE '21]{Proceedings of the 29th ACM Joint European Software Engineering Conference and Symposium on the Foundations of Software Engineering}{August 23--28, 2021}{Athens, Greece}
\acmBooktitle{Proceedings of the 29th ACM Joint European Software Engineering Conference and Symposium on the Foundations of Software Engineering (ESEC/FSE '21), August 23--28, 2021, Athens, Greece}

\usepackage{pbox}
\usepackage{fancybox}
\usepackage{graphicx}
\usepackage{float}
\usepackage{listings}
\usepackage{url} 
\usepackage{adjustbox}
\usepackage{multirow}
\usepackage{xspace}
\usepackage{comment}
\usepackage{xcolor}
\usepackage{microtype}
\usepackage{balance}
\usepackage{paralist}
\usepackage{fancyvrb}
\usepackage{algorithmic}
\usepackage{algorithm}
\usepackage{amsmath}
\usepackage{cleveref}
\usepackage{soul,listings,xcolor}
\usepackage[normalem]{ulem}
\usepackage{array}

\newcolumntype{H}{>{\setbox0=\hbox\bgroup}c<{\egroup}@{}}

\newcommand{\approach}{\textsc{Sivand}\xspace}

\lstdefinestyle{customjava}{
  belowcaptionskip=1\baselineskip,
  xleftmargin=\parindent,
  language=java,
  showstringspaces=false,
  morecomment=[s][\color{gray}]{@}{\^^M},
  keywordstyle=\bfseries\color{green!40!black},
  commentstyle=\itshape\color{purple!40!black},
  identifierstyle=\color{black},
  stringstyle=\color{orange},
}

\usepackage{color}
\definecolor{deepblue}{rgb}{0,0,0.5}
\definecolor{deepred}{rgb}{0.6,0,0}
\definecolor{deepgreen}{rgb}{0,0.5,0}
\definecolor{lightgreen}{rgb}{0.2,0.7,0.4}

\lstdefinestyle{custompython}{
  language=Python,
  morekeywords={self},              
  keywordstyle=\color{deepblue},
  stringstyle=\color{deepgreen},
  commentstyle=\color{lightgreen},
  showstringspaces=false
}

\newcolumntype{L}[1]{>{\raggedright\let\newline\\\arraybackslash\hspace{0pt}}m{#1}}
\newcolumntype{C}[1]{>{\centering\let\newline\\\arraybackslash\hspace{0pt}}m{#1}}
\newcolumntype{R}[1]{>{\raggedleft\let\newline\\\arraybackslash\hspace{0pt}}m{#1}}

\title{Understanding Neural Code Intelligence through Program Simplification}

\author{Md Rafiqul Islam Rabin}
\email{mrabin@uh.edu}
\affiliation{%
  \institution{University of Houston}
  \city{Houston}
  \state{TX}
  \country{USA}
}

\author{Vincent J. Hellendoorn}
\email{vhellendoorn@cmu.edu}
\affiliation{%
  \institution{Carnegie Mellon University}
  \city{Pittsburgh}
  \state{PA}
  \country{USA}
}

\author{Mohammad Amin Alipour}
\email{maalipou@central.uh.edu}
\affiliation{%
  \institution{University of Houston}
  \city{Houston}
  \state{TX}
  \country{USA}
}



\begin{document}

\newcommand{\Fix}[1]{\textbf{[\textcolor{red}{Fix/TODO}: #1]}}
\newcommand{\Ans}[1]{\textbf{\textcolor{blue}{Answer}: #1}}
\newcommand{\Part}[1]{\noindent\textbf{#1}}
\newcommand{\fsize}[2]{{\fontsize{#1}{0}\selectfont#2}}
\newcommand{\crossmark}{$\times$}
\newcommand{\Space}[1]{}

\newcommand{\eg}{\textit{e.g.}\xspace}
\newcommand{\ie}{\textit{i.e.}\xspace}
\newcommand{\etal}{\textit{et al.}\xspace}
\newcommand{\Fone}{\textsc{$F_1$-Score}\xspace}

\cornersize{.2}
\newcounter{observation}
\newcommand{\observation}[1]{\refstepcounter{observation}
        \begin{center}
        \vspace{3pt}
        \Ovalbox{
        \begin{minipage}{0.9\columnwidth}
            \textbf{Observation \arabic{observation}:} #1
        \end{minipage}
        }
        \end{center}
}

\newcommand{\MT}{\ensuremath{t}}
\newcommand{\OT}{\ensuremath{t_o}}
\newcommand{\RT}{\ensuremath{t_r}}

\newcommand{\vm}{\textsc{VarMisuse}\xspace}
\newcommand{\mnp}{\textsc{MethodName}\xspace}

\newcommand{\JS}{\textsc{Java-Small}\xspace}
\newcommand{\JM}{\textsc{Java-Med}\xspace}
\newcommand{\JL}{\textsc{Java-Large}\xspace}
\newcommand{\PY}{\textsc{Py150}\xspace}

\newcommand{\ctv}{\textsc{Code2Vec}\xspace}
\newcommand{\cts}{\textsc{Code2Seq}\xspace}
\newcommand{\rnn}{\textsc{RNN}\xspace}
\newcommand{\tra}{\textsc{Transformer}\xspace}

\newcounter{magicrownumbers}
\newcommand\rnum{\stepcounter{magicrownumbers}\arabic{magicrownumbers}}

    \lstnewenvironment{CODE}[1][]
      {\lstset{language=[LaTeX]TeX}\lstset{escapeinside={(*@}{@*)},
       numbers=left,numberstyle=\normalsize,stepnumber=1,numbersep=5pt,
       breaklines=true,
           framesep=5pt,
           basicstyle=\normalsize\ttfamily,
           showstringspaces=false,
           keywordstyle=\itshape\color{blue},
           stringstyle=\color{maroon},
        commentstyle=\color{black},
        rulecolor=\color{black},
        xleftmargin=0pt,
        xrightmargin=0pt,
        aboveskip=\medskipamount,
        belowskip=\medskipamount,
               backgroundcolor=\color{white}, #1
    }}
    {}

\newcounter{defn}
\newcommand{\defn}[2]{\refstepcounter{defn}
	\begin{quote}
	\textbf{#1:} #2
	\end{quote}

}

\newcommand{\OI}{\ensuremath{p_o}\xspace}
\newcommand{\RI}{\ensuremath{p_r}\xspace}
\newcommand{\M}{\ensuremath{\mathcal{M}}\xspace}
\newcommand{\MI}{\ensuremath{p}\xspace}

\newcommand{\ciinput}{input program\xspace}
\newcommand{\ciinputs}{input programs\xspace}

\newcommand{\Pred}[2]{\ensuremath{Prediction(#1, #2)}}
\newcommand{\Size}[1]{\ensuremath{|#1|}}
\newcommand{\SizeToken}[1]{\ensuremath{|#1|}$_{token}$}
\newcommand{\SizeChar}[1]{\ensuremath{|#1|}$_{char}$}
\newcommand{\DD}{delta debugging\xspace}
\newcommand{\ddmin}{\texttt{ddmin}}

\begin{abstract}
A wide range of code intelligence (CI) tools, powered by deep neural networks, have been developed recently to improve programming productivity and perform program analysis.
To reliably use such tools, developers often need to reason about the behavior of the underlying models and the factors that affect them. This is especially challenging for tools backed by deep neural networks.
Various methods have tried to reduce this opacity in the vein of ``transparent/interpretable-AI''. However, these approaches are often specific to a particular set of network architectures, even requiring access to the network's parameters. This makes them difficult to use for the average programmer, which hinders the reliable adoption of neural CI systems.
In this paper, we propose a simple, model-agnostic approach to identify critical input features for models in CI systems, by drawing on software debugging research, specifically delta debugging.
Our approach, \approach, uses simplification techniques that reduce the size of input programs of a CI model while preserving the predictions of the model. We show that this approach yields remarkably small outputs and is broadly applicable across many model architectures and problem domains. We find that the models in our experiments often rely heavily on just a few syntactic features in input programs.
We believe that \approach's extracted features may help understand neural CI systems’ predictions and learned behavior.
\end{abstract}

\maketitle

\section{Introduction}
\label{sec:intro}

Deep learning has increasingly found its use in state-of-the-art tools for code intelligence (CI), thriving in defect detection, code completion, type annotation prediction, and many more \cite{gu2016deep, white2015toward, hellendoorn2018deep}. This growing body of work benefits from the remarkable generality of deep learning models: they are \emph{universal function approximators}~\cite{hornik1989multilayer}; \ie, they can represent any function, not just linear ones. 
In practice, deep neural network methods appear to learn rich representations of raw data through a series of transforming layers~\cite{lecun2015deep}. This substantially reduces the burden of feature engineering in complex domains, including vision, natural languages, and now software.

Given this capacity, learned models seem capable of discovering many non-trivial properties about the source code, even ones that are beyond the reach of traditional, sound static analyzers.
As such, they can uncover bugs in code that is not syntactically invalid, but rather ``unnatural'' \cite{Ray:2016:NBC:2884781.2884848}.
Although this may be reminiscent of a software developer's ability to intuit properties about programs, there is a sharp contrast in interpretability: developers can explain their deductions and formulate falsifiable hypotheses about the behavior of their code.
Deep neural models offer no such capability. Rather, they remain stubbornly opaque ``black boxes'', even after years of research on interpreting their behavior \cite{Murdoch2019PNAS}.

This opacity is already a concern in non-critical applications, where the lack of explainability frustrates efforts to build useful tools. It is substantially more problematic in safety- and security-critical applications, where deep learners could play a key role in preventing defects and adversarial attacks that are hard to detect for traditional analyzers.
At present, deep neural models can change their predictions based on seemingly insignificant changes, even semantic-preserving ones \cite{rabin2019tnpa, Vechev:AdversarialCode, rabin2021generalizability, Reps:CodeRobustness}, and fail to provide any \emph{traceability} between those predictions and the code.

In this paper, we propose a simple, yet effective methodology to better analyze the input (over)reliance of neural models in software engineering applications. 
Our approach is \emph{model-agnostic}: rather than study the network itself, our approach relies on the input reductions using \DD \cite{Zeller:2002:simplifying}. 
The main insight is that by removing irrelevant parts to a prediction from the input programs, we may better understand important features in the model inference, consequently.

Given a prediction, correct or not, we show that model inputs can often be reduced tremendously, even by 90\% or more, while returning the same prediction. 
Importantly, our work is the first to show that 
these reductions in no way require the inputs to remain natural, or, depending on the task, in any way valid. 
This allows us to generate significantly simpler explanations than related work \cite{Wang:2021:demystifying, zheng2020probing}.
Our results hold on four neural architectures across two popular tasks for neural CI systems: code summarization, using Code2Vec and Code2Seq ~\cite{alon2018code2vec, alon2018code2seq}, and variable misuse detection, with RNNs and Transformers ~\cite{hellendoorn2020global}. 
We show that our minimal programs are related to, but not fully explained by the ``attention'' ~\cite{vaswani2017attention} in models that use it.
Overall, our findings suggest that current models, when trained on a single task such as ours, have very little care for the readability of functions as a whole; rather, they overwhelmingly focus on very small, simple syntactic patterns that provide salient clues to the required output.

\begin{table*}[t]
\caption{Reduction of a program while preserving the predicted method name \texttt{OnCreate} by the \ctv model.}
\begin{tabular}{c|c|Hl}
\textbf{Step(s)} & \textbf{Score} & \textbf{Loss} & \textbf{Code} 
\\ \hline \hline

1  & 1.0 &  0.0 & 
\begin{lstlisting}[style=customjava]
@Override
public void onCreate(Bundle savedInstanceState){
super.onCreate(savedInstanceState);
setContentView(R.layout.fragmentdetails);
}
\end{lstlisting}
\\ \cline{4-4}

- & - & - & 
\begin{lstlisting}[style=customjava]
...
\end{lstlisting}
\\ \cline{4-4}

4 & 1.0 & 0.0  &
\begin{lstlisting}[style=customjava]
void onCreate(Bundle savceState) { super.onCreate(snceState); 
setContentView(R.layout.fragmentdetails); }
\end{lstlisting}
\\ \cline{4-4}

7 & 1.0 & 0.0  &
\begin{lstlisting}[style=customjava]
void onCreate(Bundle savceState) { super.onCreate(snceState); setContentView(s); }
\end{lstlisting}
\\ \cline{4-4}

10 & 0.99 & 0.0001  &
\begin{lstlisting}[style=customjava]
void onCreate(Bu savate) { s.onCreate(snceState); setContentView(s); }
\end{lstlisting}
\\ \cline{4-4}

13 & 0.98 & 0.02  &
\begin{lstlisting}[style=customjava]
void onCreate(Bu savate) { s.onCreate(snte); View(s); }
\end{lstlisting}
\\ \cline{4-4}

15 & 0.62 & 0.52  &
\begin{lstlisting}[style=customjava]
void onCreate(Bu savate) { s.onCreate(snte);(s); }
\end{lstlisting}
\\ \cline{4-4}

18 & 0.46 & 0.84  &
\begin{lstlisting}[style=customjava]
id onCreate(Bu save) { s.onCreate(snte);; }
\end{lstlisting}
\\ \cline{4-4}

27 & 0.86 & 0.19  &
\begin{lstlisting}[style=customjava]
d onCreate(u ve){s.onCreate(e); }
\end{lstlisting}
\\ \cline{4-4}

- & - & - & 
\begin{lstlisting}[style=customjava]
...
\end{lstlisting}
\\ \cline{4-4}

29 & 0.79 & 0.30 &
\begin{lstlisting}[style=customjava]
d onCreate(u ve){s.onCreate();}
\end{lstlisting} 
\\ \hline

\end{tabular}
\label{tab:dd-example-cts}
\end{table*}

\begin{figure}[t]
\textbf{Original}
\begin{lstlisting}[style=customjava]
@Override
public void onCreate(Bundle savedInstanceState){
    super.onCreate(savedInstanceState);
    setContentView(R.layout.fragmentdetails);
}
\end{lstlisting}
~\\
\textbf{Minimized}
\begin{lstlisting}[style=customjava]
d onCreate(u ve){s.onCreate();}
\end{lstlisting}
\caption{Example of an original and minimized method in which the target is to predict \lstinline{onCreate} as the method name. The minimized example clearly shows that the model (i.e. \ctv) has learned to take \underline{short-cuts}, in this case looking for the name in the function's body.}
\label{lst:oncreate-example}
\end{figure}

\section{Motivating Example}
\label{sec:motivation}
We first present an example to illustrate our approach and the insights that it can provide about the prediction of neural code intelligence models. \Cref{lst:oncreate-example} shows a code snippet from the \mnp dataset. Here, the goal is to infer the method name from the method body. The \ctv model predicts the method name \texttt{onCreate}, which is indeed correct. Unfortunately, it is hardly clear why; \ctv considers dozens to hundreds of ``paths'' in the methods' AST (abstract syntax tree) and struggles to identify which ones are most informative -- indeed, in our results, we find its built-in ``attention'' based explanations to only poorly correlate with essential tokens.

Of course, a developer looking at this program could provide several explanations why this is intuitively correct. For one, the method invocation \lstinline{super.onCreate} strongly suggests that this method overrides a method with the same name in the parent class, as such calls are usually made to the overridden method.
This guess does not need to be informed by just that call; if the developer is familiar with Android development, they might recognize this code as working with Android APIs, and perhaps even know that the \texttt{Activity}\footnote{\url{https://developer.android.com/reference/android/app/Activity}} in Android systems provides \texttt{onCreate} method to initialize a user interface in which developers can place their UI objects on the interface window using the \texttt{setContentView} method.

Turning back to the \ctv model, it does not offer any of these explanations, but rather an output based on a complex mixture of its inputs and millions of parameters. This lack of transparency is problematic for practitioners and researchers alike, who are unlikely to accept recommendations without \emph{evidence}, especially in cases that are not so trivial. For instance, does \ctv note and use the \lstinline{.onCreate} invocation? If so, does it specifically leverage the inheritance relation? And/or, does it use any of the more Android-specific reasoning about the presence of the \texttt{setContentView} invocation?

Our approach, \approach, can better provide compelling answers to these questions. The second half of \Cref{lst:oncreate-example} shows the smallest possible version of this program that is still syntactically valid (a necessity for \ctv) and yields the same prediction. Any evidence of this method relating to Android development has been thoroughly scrubbed; all that remains is the single \lstinline{onCreate} method call -- even the mention of \lstinline{super} has been minified to a single character `s'.

\Cref{tab:dd-example-cts} shows some steps that yielded this reduced program in greater detail.
\approach works by iteratively reducing the size of the input program by the \ctv model while preserving the prediction output at every step.
We continue this reduction until the program is either fully reduced (to its minimal components, depending on the task) or any further reduction would corrupt the prediction.
Each row in this table shows the intermediate, most reduced program that still yields the \emph{same} prediction as the original input program in the first row, as well as the probability (``score") of that prediction for reference. A close examination of the intermediate results suggests that
the \lstinline{savedInstanceState} parameter and the \lstinline{setContentView} call, indicative of Android development, is reduced with almost no penalty to the score. Other such cues follow soon after. In step 10, the \lstinline{super} call, which represents cues of inheritance, is similarly pruned, and here too with virtually no penalty to the score. The subsequent steps largely serve to remove a few miscellaneous characters; these compromise the prediction score somewhat, perhaps because the method is increasingly irregular. Regardless, as long as \lstinline{onCreate} is present, the score remains high, ample to sustain the prediction.
Evidently, even if the model noticed the Android-related features, it certainly did not need them, nor even the mention of \lstinline{super}; all that remains is the presence of another method call \lstinline{onCreate}.

That \approach can elucidate these insights is a mixed blessing and curse: its model-agnostic approach effectively lets us bypass interpreting the millions of parameters and complex inner workings of the studied models, and yields remarkably short programs that make the original predictions easier to comprehend. The use of \DD \cite{Zeller:2002:simplifying} is a novel approach to such model interpretation in general, and its intermediate steps evidently provide useful insights into the model's ``thinking''. 
At the same time, our approach is a notable departure from more common neural attribution methods (\eg \cite{Attribution}), which typically try to find the part of the input that was most informative, but do not necessarily assume that all other parts could be removed entirely.
\approach, on the other hand, frequently corrupts its input programs almost beyond recognition. That the assessed models continue to perform so well during this process (\Cref{sec:results}), across multiple code intelligence tasks, is indicative of these models' overreliance on small features and lack of holistic interpretation of their input programs.
We discuss various interpretations and implications of this phenomenon in \Cref{sec:discussion}.

\section{\approach Methodology}
\label{sec:approach}
This section describes \approach, a methodology for better understanding and evaluating CI models. This methodology is model-agnostic and can be used for any CI model, however, in this paper, we will focus on the CI models that use neural networks for training, as their lack of transparency warrants more attention.

\subsection{Notation}
We use \OI{} to denote the original input to the neural CI model and \RI{} to denote the simplified input.
Let \M{} be an arbitrary CI model, and \MI an arbitrary input program, then \Pred{\M}{\MI} denotes the prediction result of \M given \MI, and \Size{\MI} the resultant size of \MI.
Conceptually, we define size as the number of atomic parts that \MI has. In the evaluation of this approach, we use token as the atomic part, hence \Size{\MI} denotes the number of tokens that are returned by the lexer of the language.

The high-level goal of \approach is to produce a reduced \ciinput \RI that is smaller than the size of \OI, i.e.,
$\Size{\RI} < \Size{\OI}$ (and ideally 
$\Size{\RI}\ll\Size{\OI}$),
such that if $\Pred{\M}{\OI} = r$, \RI{} still retains the same prediction $r$: $\Pred{M}{\RI} = r$. 
While such a reduction process can in principle terminate under many kinds of conditions (and even return the original \ciinput), we are interested in finding the so-called ``1-minimal'' \ciinput~\cite{Zeller:2002:simplifying}, where no single part of \RI can be removed without losing some desired property of \OI.

\subsection{Workflow in \approach}
\begin{figure*}
    \centering
    \includegraphics[width=0.8\linewidth]{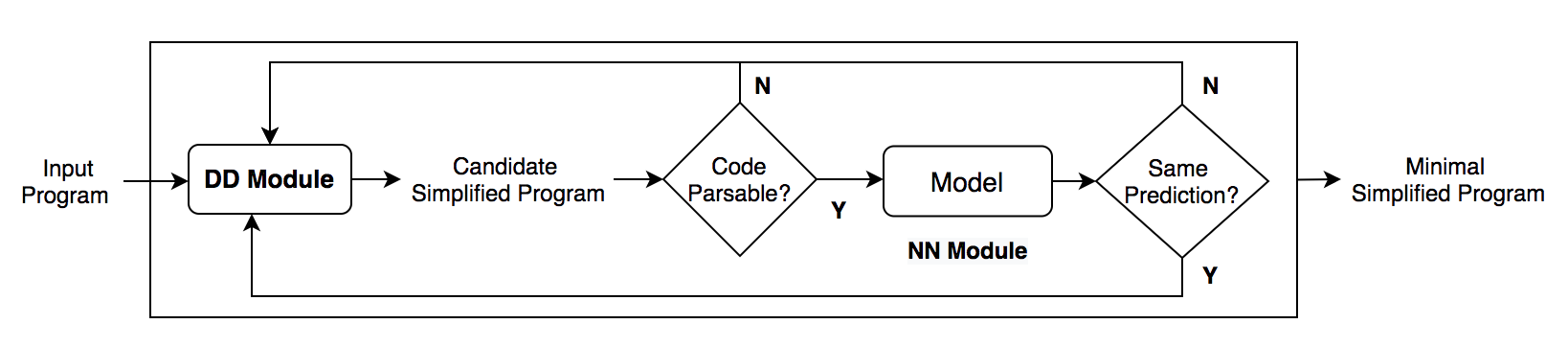}
    \caption{Workflow of \approach.}
    \label{fig:approach}
    
\end{figure*}

Figure ~\ref{fig:approach} depicts a high-level view of the workflow in the proposed methodology, \ie \approach. Given a \ciinput, \approach uses the DD module to reduce the size of the program. The DD module uses \DD to produce various candidate programs by removing various parts of the original \ciinput
and iteratively searches for the 1-minimal \ciinput that produces the same prediction as the original \ciinput.
Some of the generated candidates might be invalid programs; that is, they do not follow the syntax of the programming language that the program is written in. Therefore, since some CI models\Space{, such as \ctv~\cite{alon2018code2vec} and \cts~\cite{alon2018code2seq},} require inputs to be syntactically valid; to enhance the efficiency, \approach filters out the invalid candidates that do not parse only for those CIs.

\subsection{Reduction Algorithm}
\label{sec:algorithm}


\renewcommand{\algorithmicrequire}{\textbf{Input:}}
\begin{algorithm}
 \caption{High level algorithm for  $ddmin$ \DD. The algorithm is initiated by $ddmin(\M, \OI, 2)$.}
 \label{alg:ddmin}
  \begin{algorithmic}
    \REQUIRE \M, CI model; \MI, \ciinput; and $n$, number of partitions.
    \STATE Split \MI~into $n$ partitions to build $\Delta_1, ..., \Delta_n$
    \IF {$\exists \Delta_i$ such that $\Pred{\M}{\OI} == \Pred{\M}{\Delta_i}$}
    \STATE $ddmin(\M,~\Delta_i,~2)$
    \ELSIF{$\exists \Delta_i$ s.t.  $\Pred{\M}{\OI} == \Pred{\M}{\MI - \Delta_i}$ }
    \STATE $ddmin(\M,~$\MI-$\Delta_i,~max (n-1, 2))$
    \ELSIF{$n < \Size{\MI}$ }
    \STATE $ddmin(\M,~\MI,~min (2n, \Size{\MI}))$
    \ELSE
    \STATE return \MI
    \ENDIF
 \end{algorithmic}
\end{algorithm}

Algorithm ~\ref{alg:ddmin} describes the \DD algorithm proposed by Zeller and Hildebrandt~\cite{Zeller:2002:simplifying} and later extended by Groce \etal~\cite{Groce:2014:CauseReduction:ICST, Groce:2016:CauseReduction:STVR,Alipour:2016:NonAdequate:ASE}, that is adapted to our task for finding  minimal \ciinputs.
At a high level, the \DD algorithm iteratively splits a \ciinput into multiple candidate programs by removing parts of the \ciinput. It uses $n$ to specify the granularity of parts. That is, for an \ciinput \MI and granularity level $n$, it generates $2n$ candidates: $n$ candidates by splitting \MI into $n$ partitions $\Delta_1,\ldots, \Delta_n$, and another $n$ candidates by computing $\MI-\Delta_1,\ldots, \MI-\Delta_n$.
At each of these steps, the algorithm checks if any resulting candidate program \MI satisfies the desired property, which here means preserving the prediction of the model on the original \ciinput, i.e., $\Pred{\M}{\OI} = \Pred{\M}{\Delta}$, where \OI denotes the original \ciinput.
When the algorithm finds a candidate satisfying the property, it uses the candidate as the new base \MI to be reduced further; otherwise, the algorithm increases the granularity, i.e., $n$ for splitting, until the algorithm determines that the \MI is 1-minimal.  
The time complexity of the \DD algorithm is quadratic, i.e. $O(n^2)$ in the size of the input program.

\section{Experimental Settings}
\label{sec:settings}
Our proposed methodology is task- and program agnostic. We assess these properties by studying two tasks, and two models on each of these. This section outlines all of these.

\subsection{Subject Tasks}
We study two popular code intelligence tasks that have gained interest recently: method name prediction (\mnp) \cite{allamanis2016summarization, alon2018code2vec, alon2018code2seq}, and variable misuse detection \vm \cite{allamanis2018learning,vasic2019neural,hellendoorn2020global}.

\subsubsection{\mnp}
In the method name prediction task, the model attempts to predict the name of a method, given its body.
We study two commonly used, and similar, neural approaches for this task: \ctv~\cite{alon2018code2vec}, and \cts~\cite{alon2018code2seq}. Both these approaches rely on extracting ``paths'' in the method's AST, that connect one terminal (token) to another, which are mapped to vector embeddings. 
These paths are enumerated exhaustively and used by the two models in various ways. The premise is that these paths consolidate both lexical and syntactic information, thus providing more benefit than strictly token-based models, such as RNNs.

In \ctv~\cite{alon2018code2vec}, each path, along with its source and destination terminals, is mapped into a vector embedding that is learned jointly with other network parameters during training.
The separate vectors of each path-context are then concatenated to a single context vector using a fully connected layer.
An attention vector is also learned with the network, which is used to aggregate the path-context representations into a single code vector representing a method body.
The model then predicts the probability of each target method name given the code vector of the method body via a softmax-normalization between the code vector and each of the embeddings of a large vocabulary of possible method names.

The \cts~\cite{alon2018code2seq} model uses an encoder-decoder architecture to encode paths node-by-node and generate labels as sequences at each step.
In \cts, the encoder represents a method body as a set of paths in AST where individual paths are compressed to a fixed-length vector using a bi-directional LSTM, which encodes paths node-by-node with splitting tokens into sub-tokens.
The decoder again uses attention to select relevant paths while decoding, and predicts sub-tokens of a target sequence at each step when generating the method name.

\subsubsection{\vm}
A variable misuse bug \cite{allamanis2018learning} occurs when the intended variable used at a particular location of the program differs from the actual variable used. These mistakes are commonly found in real software development \cite{karampatsis2020often}, and naturally occur as ``copy-paste bugs''. We specifically experiment with the joint bug localization and repair task of \vm \cite{vasic2019neural}: given a tokenized sample (buggy or correct), the task is to predict two pointers into these tokens: a) localization: one pointer to indicate the location of the token that contains the wrong variable, or some default token if no bug exists, and b) repair: another pointer to indicate the location of any token that contains the correct variable (ignored for bug-free samples).

We specifically use the dataset released by Hellendoorn \etal \cite{hellendoorn2020global}, whose repository also includes a number of models that be applied directly to this dataset.\footnote{\url{https://github.com/VHellendoorn/ICLR20-Great}} From these, we use the following two generic neural approaches: \rnn, and \tra.
The \rnn model here is a simple bi-directional recurrent architecture that uses GRU as the recurrent cell, and has 2 layers, and 512 hidden dimensions.
The \tra model is based entirely on attention, in which the representations of a snippet's tokens are repeatedly improved through combination with those of all others in the functions. We use the parameters from the original Transformer \cite{vaswani2017attention}, with 6 layers, 8 heads, and 512 attention dimensions.

\subsection{Datasets and Models}
\label{sec:dataset}

Table~\ref{table:all_models} summarizes the performance characteristics of the CI models that we use in our experiments which is on par with the ones reported in the original studies. 

For the \mnp task, we use the \JL dataset\footnote{\url{https://github.com/tech-srl/code2seq\#java}} to train both the \ctv and \cts models. This dataset contains a total of $9,500$ Java projects from GitHub, partitioned into $9,000$ projects as training data, $200$ projects as validation data, and $300$ projects to test on. Overall, it contains about $16$M methods.

For the \vm task, we use the \PY corpus \footnote{\url{https://github.com/VHellendoorn/ICLR20-Great\#data}} to train both the \rnn and \tra models. This dataset contains functions from a total of $150$K Python files from GitHub, and are already partitioned into $90$K files as training set, $10$K files as validation set, and $50$K files as testing set. Each function is included both as a bug-free sample, and with up to three synthetically introduced bugs, yielding about 2 million samples in total.

\begin{table}
    \begin{center}
        \caption{Characteristics of the trained models.}
        \label{table:all_models}
        \resizebox{\columnwidth}{!}{%
        \begin{tabular}{|c|c|C{2cm}|C{2cm}|C{2cm}|}
            \hline
            \textbf{Model} & \textbf{Dataset} & 
            \textbf{Precision} & \textbf{Recall} & \textbf{$F_1$-Score} \\ \hline 
            \hline
            \ctv & \JL & 45.17 & 32.28 & 37.65 \\ \hline 
            \cts & \JL & 64.03 & 55.02 & 59.19 \\ \hline 
            \hline
            
            \textbf{Model} & \textbf{Dataset} & 
            \textbf{Localization Accuracy} & \textbf{Repair Accuracy} & \textbf{Joint Accuracy} \\ \hline 
            \hline
            \rnn & \PY & 63.56 & 63.22 & 52.18 \\ \hline
            \tra & \PY & 73.39 & 76.79 & 66.05 \\ \hline
        \end{tabular}%
        }
    \end{center}
\end{table}

\subsection{Sample Inputs}

To evaluate the effectiveness of \approach, we choose both correctly and incorrectly predicted samples from the test set of these datasets.

\noindent a) Correctly predicted samples:
\begin{itemize}
    \item For \mnp task, we choose 1000 correctly predicted samples for token-level reduction, where the smallest unit of reduction is a token, and around 500 correctly predicted samples for character-level reduction, where the smallest unit is a character. Running character-based reduction was slow as the search space of possible reductions is much larger, hence the lower total. We use the same randomly selected samples for both \ctv and \cts models.
    \item For \vm task, we choose 2000 correctly predicted samples: 1000 from buggy samples, and 1000 bug-free ones. For the selected buggy samples, models correctly predicted both the location pointer and repair targets. For the selected non-buggy samples, models correctly identify as no-bug by prediction special 0-index. We use the same randomly selected samples for both \rnn and \tra models, thus ensuring that we selected only samples on which their predictions were both correct.
\end{itemize}

\noindent b) Wrongly predicted samples:
\begin{itemize}
    \item For \mnp task, we choose around 500 samples where the predicted method name is wrong. We use token-level reduction only, and randomly select different wrong samples for \ctv and \cts, separately.
    \item For \vm task, we choose 1000 buggy samples where the predicted location pointer is wrong but models correctly predict the repair targets. We randomly select different wrong samples for \rnn and \tra, separately.
\end{itemize}

\subsection{Metrics}
Here, we define the metrics that we measure in the experiments as follows. 

We use \emph{size reduction ratio} of \ciinputs to evaluate the effectiveness of \approach in reducing the size of the original \ciinputs.
For the \ciinput \OI, and its reduced counterpart \RI, size reduction ratio (or Reduction\%) is calculated as $100*\frac{\Size{\OI}-\Size{\RI}}{\Size{\OI}}$.

All neural network models in our work are capable of making probabilistic predictions, in which a probability distribution is inferred over either tokens (for \vm) or method name (parts) (for \mnp). We can leverage these prediction probabilities to compute a ``prediction score'', that indicates the confidence of the model in a particular prediction. When we reduce the inputs, we track changes to these scores on the original targets; the reduction is stopped once the model ceases to predict the correct output, which generally comes with a drop in the score (probability) of that target. To assess whether our models lose certainty during this reduction phase, we study the changes in the distribution of these scores relative to program reductions.

\paragraph{Hardware}
We used a server with an Intel(R) Xeon(R) 2.30GHz CPU and a single NVIDIA Tesla P100 GPU with 12GB of memory to run the experiments in this study.

\paragraph{\textbf{Artifacts}}
The code and data of \approach are publicly available at \url{https://github.com/UH-SERG/SIVAND}.

\section{Results}
\label{sec:results}
In this section, we present the results of our experiments where we seek to answer the following research questions.
\begin{itemize}
    \item[RQ1] How much can typical input programs be reduced?
    \item[RQ2] What factors influence reduction potential?
    \item[RQ3] Do reduced programs preserve the tokens most used by attention mechanisms?
    \item[RQ4] What is the cost of running \approach? 
\end{itemize}

Note that for brevity, in this section, unless otherwise noted, the results belong to token-level reduction of correct predictions, and in \vm models, buggy programs.

\subsection{RQ1: Length Reduction}

\begin{figure}[t]
    \centering
    \includegraphics[width=\columnwidth]{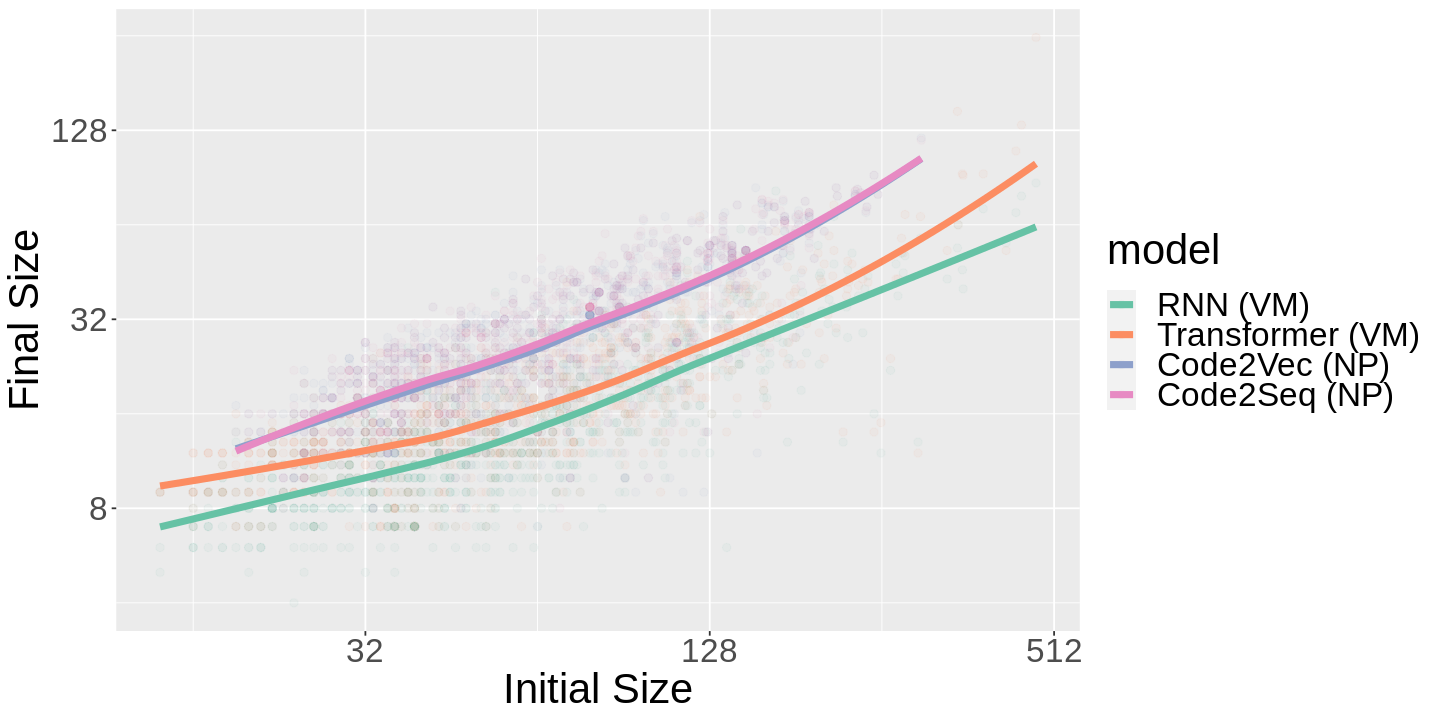}
    \caption{Initial vs. final size of reduced programs in our dataset, measured in tokens. Note the log-scaling on both axes.}
    \label{fig:initsize-vs-final-absolute}
\end{figure}

The goal of \approach is to reduce the size of programs as much as possible to help uncover features that highly impact the prediction.
~\Cref{fig:initsize-vs-final-absolute} shows its capacity to do so in terms of the size of the original programs versus the reduced programs, measured in tokens. This plot aggregates the results of 1,000 such reductions in which each program is reduced until just before its prediction changes. In this, and similar figures, we show the LOESS trend to amortize that large volume of data points (also shown with low opacity). The confidence intervals on these trends are generally very tight and mostly omitted, except where informative.

\begin{figure}[t]
    \centering
    \includegraphics[width=\columnwidth]{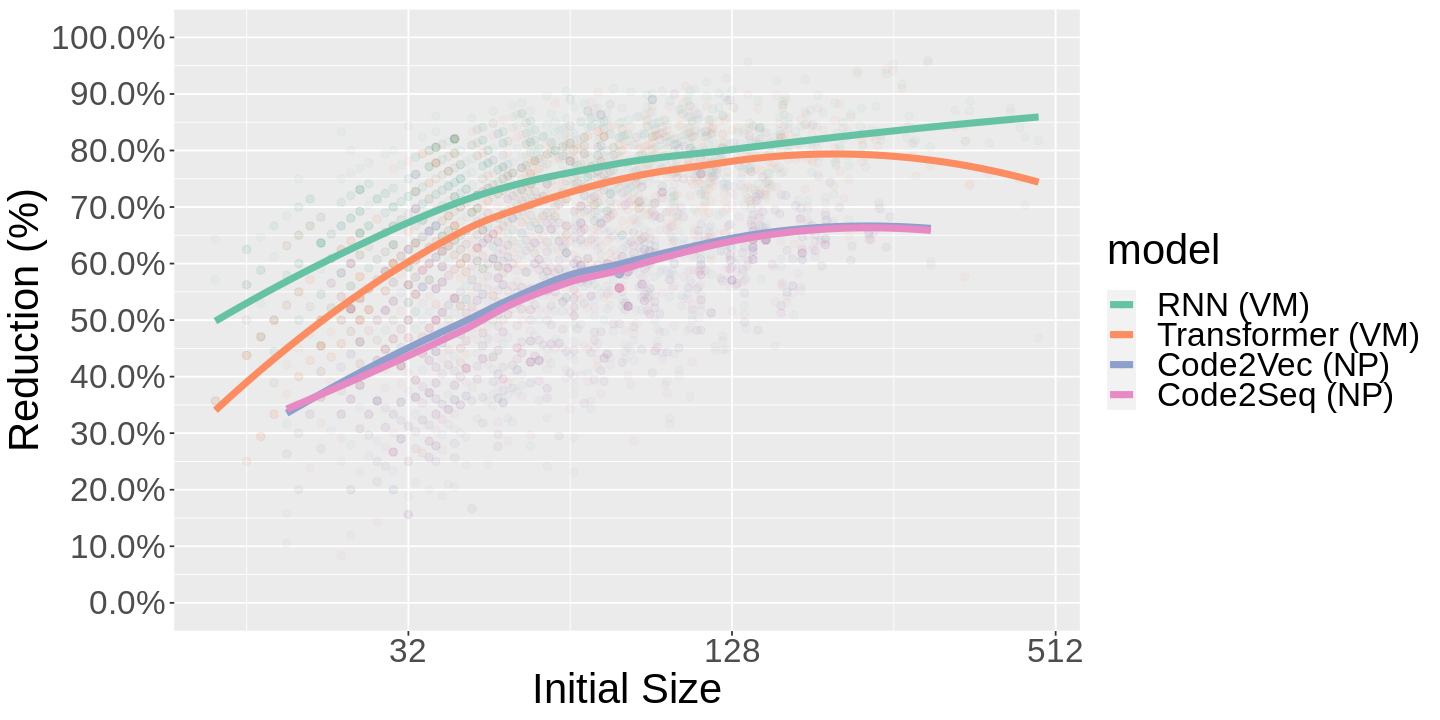}
    \caption{Percentage of tokens reduced vs. initial program size in the various models studied. Note the log-scaling on the x-axis.}
    \label{fig:percent-reduction}
\end{figure}

We find this relation to be mostly linear across all our datasets and models, with the final program containing 2 to 5 times fewer tokens than the initial one. \Cref{fig:percent-reduction} provides an alternative view, showing the achieved reduction as a percentage of the input program size; evidently, program reduction is somewhat easier on larger programs, perhaps because those are more redundant.

\begin{figure}[t]
    \centering
    \includegraphics[width=\columnwidth]{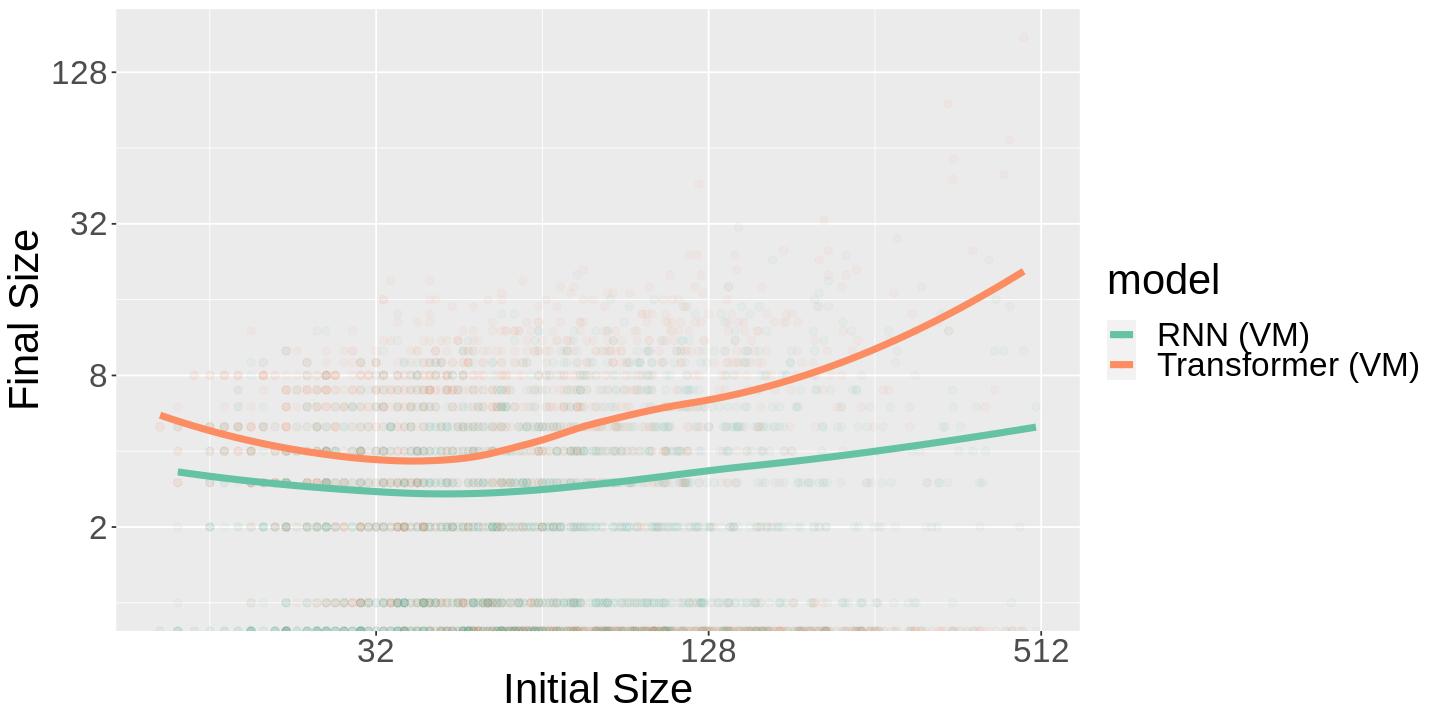}
    \caption{Tokens remaining relative to the minimum possible reduction, in the \vm task. Note the log-scaling on both axes.}
    \label{fig:rel-reduction}
\end{figure}

These figures alone suggest that the \approach can reduce a large portion of the \ciinputs, but we found that the true reduction is more substantial: in both datasets, the \emph{maximum possible reduction} is limited, in the \vm case by the need to preserve all variable occurrences (which are error and repair targets), and in the \mnp case by requiring the program to be syntactically valid (which requires keeping at least some syntactic tokens). The true minimum is less tractable for the latter dataset, but when we subtract the irreducible portion from the former, we found the relation shown in \Cref{fig:rel-reduction}. Interestingly, for most programs, this trend was \emph{nearly constant}, especially for the \rnn, which on average preserves just a few ``extra'' tokens. The Transformer, meanwhile, appears to be especially reliant on additional syntax on larger programs, although still just a fraction of the original (some 5\% of non-essential tokens). In other words, both models require nearly no additional syntax than just the variable names to make their predictions.

The ``Reduction(\%)'' columns in Table~\ref{tbl:avg_summary} provides details of the ratio of reduced tokens to the number of maximum allowed tokens in the models.
It shows that our \approach could remove more than 10\% of tokens from each input programs, on average, it removes $62.39\%$, $61.22\%$, $93.61\%$, and $89.43\%$ of the maximum allowed tokens of the original inputs in \ctv, \cts, \rnn, and \tra models, respectively.
Also in some cases, \approach was able to remove \emph{all} or almost all of the maximum allowed tokens and isolate the important features.
This substantial reduction allows to better understand and pinpoint important features in the prediction of the models.

\observation{\approach can reduce the input programs significantly: on average, more than $61\%$ in \mnp models, and more than $89\%$ in \vm models. Different neural architectures show slightly different behaviors, with \tra permitting less reduction than \rnn.}

\subsection{RQ2: Factors Impacting Reduction}
We next study a range of factors that impact, and are impacted by our program reduction approach, to better understand its effect.

\begin{figure}
    \centering
    \includegraphics[width=0.9\columnwidth]{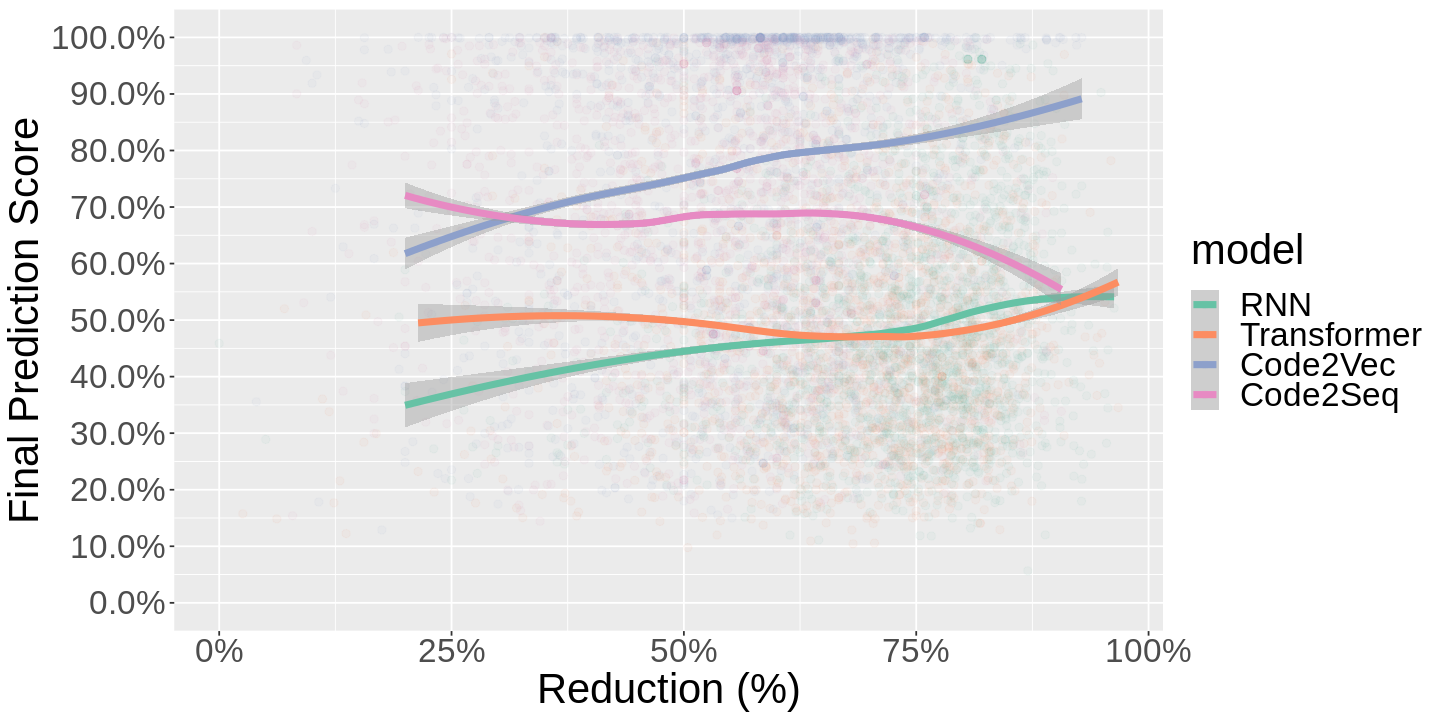}
    \caption{The final score of maximally reduced programs (immediately before changing predictions) vs. the degree of reduction.}
    \label{fig:reduction-prediction-score}
\end{figure}

\paragraph{Impact on prediction score:}
\Cref{fig:reduction-prediction-score} tracks the final prediction score against the fraction of the program that was reduced, with inter-quartile ranges shaded (``buggy'' samples only for \vm).
Recall that the reduction process halts when any further reduction would change its prediction, so it is expected that scores remain modestly high (although not necessarily above 50\%). The \ctv and \cts models, and to a lesser extent the \tra model, display pronounced ``tipping point'' behavior, in which the final reduction step still preserved relatively high probabilities while the immediate next step would have to drop at least below 50\%. 
Note that most samples started out with a score of almost exactly 100\%, regardless of models; thus, the difference  between the initial and final score is not especially informative.
The degree of reduction does not appear to impact the prediction scores much; only for the \rnn model on the \vm task do we see a slight downwards trend among input programs that could not be reduced by much, usually smaller ones. We found similarly little correlation with the input and final reduced program sizes. This further reinforces that just one or a few tokens make the difference between a confident prediction and a misprediction.

\begin{figure}
    \centering
    \includegraphics[width=0.95\columnwidth]{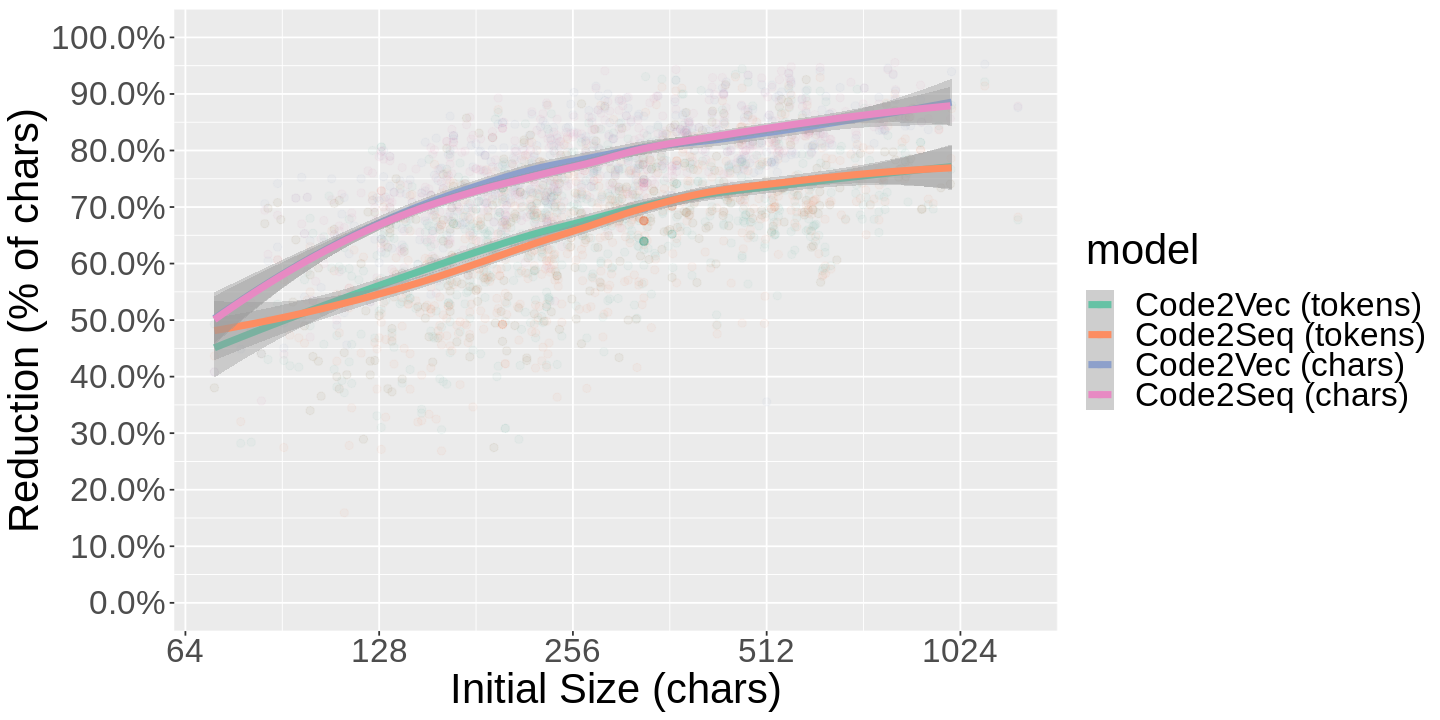}
    \caption{Character vs. token-based model reduction potential on the \mnp task, both compared in terms of total characters reduced. Note the log-scaling on the x-axis.}
    \label{fig:chars_vs_tokens}
\end{figure}

\paragraph{Character vs. token level:}
On the \mnp dataset, we additionally studied character-level \DD (besides the token-level default). This has the potential to reduce inputs beyond what is possible with token level models, and \Cref{fig:chars_vs_tokens} confirms that it frequently does: this based approach is able to remove another 10-20\% of the characters in the input program, thus yielding shorter, and potentially more informative reductions (such as the one in \Cref{lst:oncreate-example}).

\paragraph{Incorrect predictions:}
Finally, we analyzed whether models could reduce evidence for \emph{incorrect} predictions as well: so far, all our analysis focused on correctly predicted samples, but of course, in practice all these models regularly make mistakes. We can extract the reduced programs for these mispredictions in much the same way, by setting the goal for \approach to preserve the incorrect target while reducing the program. Here, we found basically no effect: the correct and incorrect predictions could on average (across all models) be reduced by 62.9\% and 64.8\% respectively\Space{, a significant but very minor difference}. Similarly, this did not seem to correlate much with the initial program size for any of our models. Thus, it appears that the models do not differentiate in terms of the evidence required to mispredict: correct or not, the models base their prediction on small parts of the input program.

\begin{table*}
    \caption{Summary of token reduction results on the correctly predicted samples.}
    \label{tbl:avg_summary}
    \vspace{-2mm}
    \def\arraystretch{1.1}
    \resizebox{0.98\textwidth}{!}{%
    \begin{tabular}{|c|c|Hc|c|c|c|c|c|c|c|c|c|c|c|c|c|c|c|}
        \hline
        \multirow{2}{*}{\textbf{Task}}
        & \multirow{2}{*}{\textbf{Model}}
        & \multirow{2}{*}{\textbf{Dataset}}
        & \multirow{2}{*}{\textbf{\# Sample}}
        & \multicolumn{3}{c|}{\textbf{\# Tokens}}
        & \multicolumn{3}{c|}{\textbf{Reduction (\%)}}
        & \multicolumn{3}{c|}{\textbf{Common (\%)}}
        & \multicolumn{3}{c|}{\textbf{\# DD Pass (Average)}} 
        & \multicolumn{3}{c|}{\textbf{Time (second)}} \\ \cline{5-19}
        
        & & & 
        & \textbf{Min} & \textbf{Avg} & \textbf{Max}
        & \textbf{Min} & \textbf{Avg} & \textbf{Max}
        & \textbf{Min} & \textbf{Avg} & \textbf{Max}
        & \textbf{Total} & \textbf{Valid} & \textbf{Correct} 
        & \textbf{Min} & \textbf{Avg} & \textbf{Max} \\ 
        
        \hline \hline
         \multirow{2}{*}{\mnp}
         & \ctv & \multirow{2}{*}{\JL} & \multirow{2}{*}{1000} & 19 & 76.08 & 300 & 15.38 & 62.39 & 98.28 & 21.43 & 63.97 & 93.33 & 351.22 & 37.01 & 33.75 & 29.44 & 106.32 & 714.56 \\ \cline{2-2}\cline{5-19}
         & \cts &  &  & 19 & 76.08 & 300 & 11.11 & 61.22 & 96.88 & 36.84 & 72.57 & 95.00 & 355.01 & 37.60 & 33.60 & 13.99 & 59.48 & 216.14\\
        \hline\hline  

        \multirow{2}{*}{\vm}
         & \rnn & \multirow{2}{*}{\PY} & \multirow{2}{*}{2000} & 13 & 82.41 & 501 & 29.41 & 93.61 & 100 & - & - & - & 237.32 & 43.53 & 30.57 & 0.13 & 3.68 & 77.57 \\ \cline{2-2}\cline{5-19}
         & \tra &  &  & 13 & 82.41 & 501 & 20.00 & 89.43 & 100 & 41.67 & 85.85 & 100 & 241.48 & 47.01 & 29.57 & 0.11 & 2.99 & 108.67\\
        
        \hline
        
        \end{tabular}%
        }
\end{table*}

\subsection{RQ3: Similarity Between Attention \& Reduction}

Attention mechanism \cite{vaswani2017attention} in neural networks attempts to capture the important features in the input and directly connect them to the output layer to increase the impact in the prediction and improve the performance of neural networks and accelerate training. Intuitively, attention captures the important features in the \ciinputs; therefore, the number of features shared between \approach and attention can indicate the effectiveness of the approach for retaining important features.

To evaluate the similarity between tokens in reductions and attention, we apply the \approach to reduce the \ciinput \OI to \RI. Let $T_{dd}$ denotes the set of tokens in $\RI$, and $k$ is the length of $T_{dd}$. To evaluate how much the set of features retained by \approach overlap with the features used in the attention mechanism, for each \OI, we then collect the set of tokens, $T_{attn}$ in the original \ciinputs that receives high attention score. For a fair comparison, we select almost the same size of attention tokens as in the reduced program. From \tra model, we get the attention score for each token, hence, we select top-$k$ tokens as $T_{attn}$. However, for the \ctv and \cts models, we get an attention score for each path between two terminal nodes in the program's AST. Therefore, we instead repeatedly choose high attention paths and collect all nodes until we found $k$ distinct nodes. Finally, we calculate the similarity between the tokens in the reduced \ciinput and the attention mechanism by computing
$\frac{|T_{attn} \cap T_{dd}|}{|T_{dd|}}$.

\begin{figure}
    \centering
    \includegraphics[scale=0.4,trim=20 40 20 30 ]{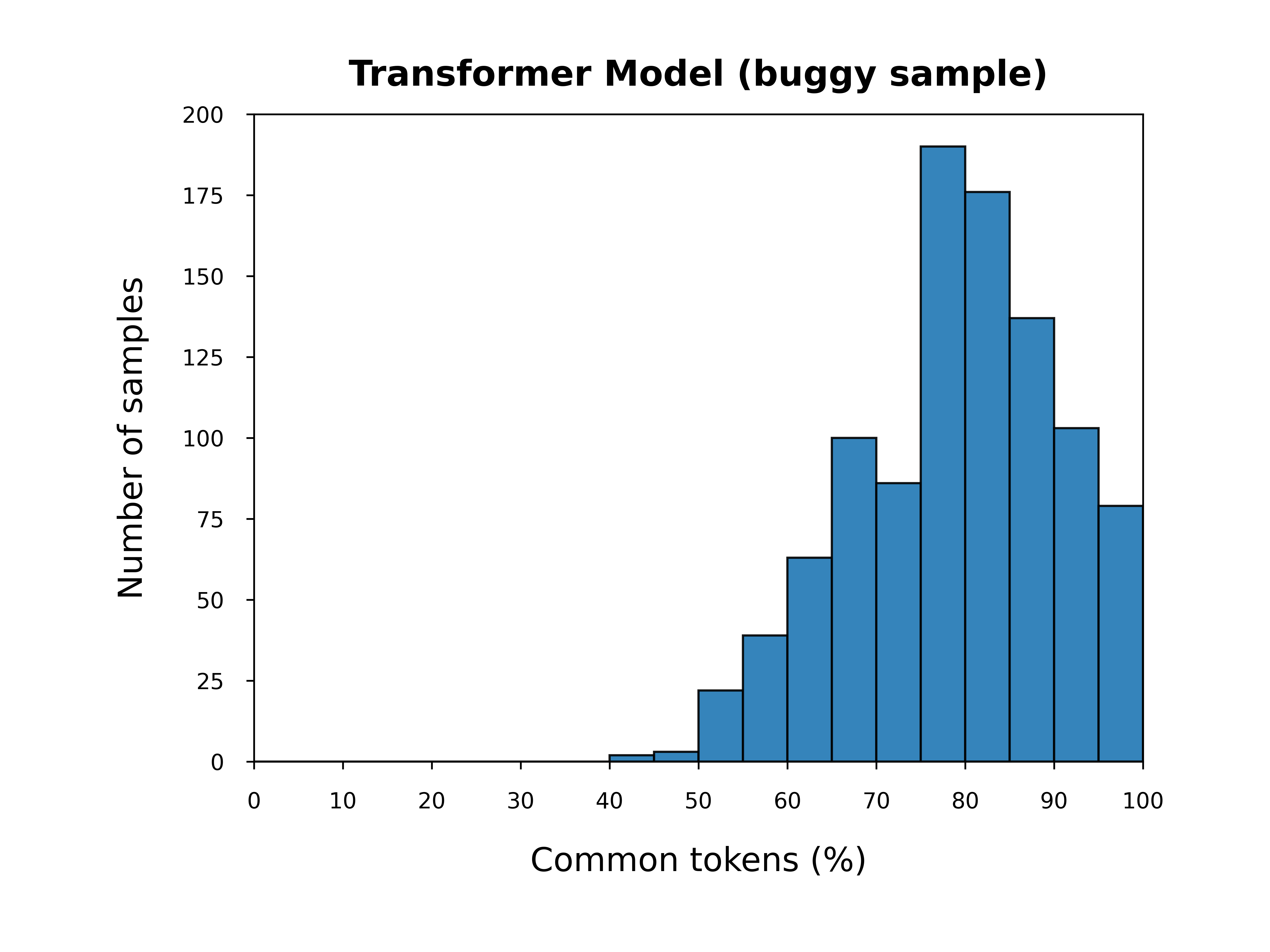}
    \caption{Percentage of common tokens between attention and reduced input programs in \tra for buggy inputs.}
    \label{fig:tra:buggy:match}
\end{figure}

\begin{figure}
        \includegraphics[scale=0.4,trim=20 40 20 30]{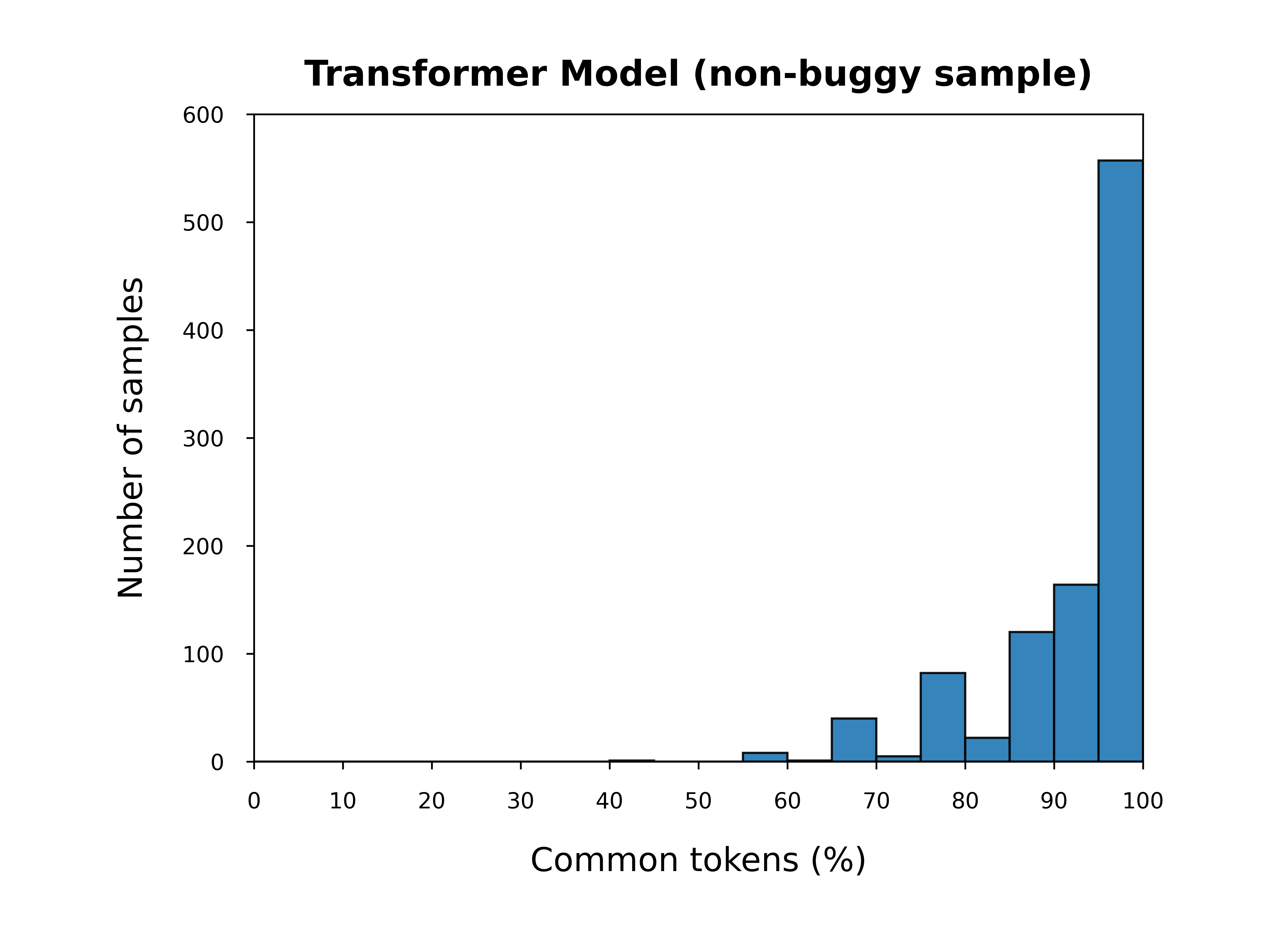}
        \caption{Percentage of common tokens between attention and reduced input programs in \tra for non-buggy inputs.}
       \label{fig:tra:non-buggy:match}
\end{figure}

Figures~\ref{fig:tra:buggy:match} and~\ref{fig:tra:non-buggy:match} show the similarity between the tokens in the attention and reduction in \tra model.
The results suggest that in the majority (in more than 550 cases) of the non-buggy \ciinputs $T_{attn}$ and $T_{dd}$ fully match, and in buggy \ciinputs there is a large overlap, majority over $80\%$ match. According to the column `Common(\%)' in Table~\ref{tbl:avg_summary}, we can see that the \ctv and \cts also have an average overlap over 60\% and 70\%, respectively. Note that \rnn in this experiment does not use attention.

\observation{On average, a large portion of tokens (between 60\% and 80\%) are shared between attention tokens and tokens retained by the reduction.}

\begin{figure}
    \centering
    \includegraphics[width=0.95\columnwidth]{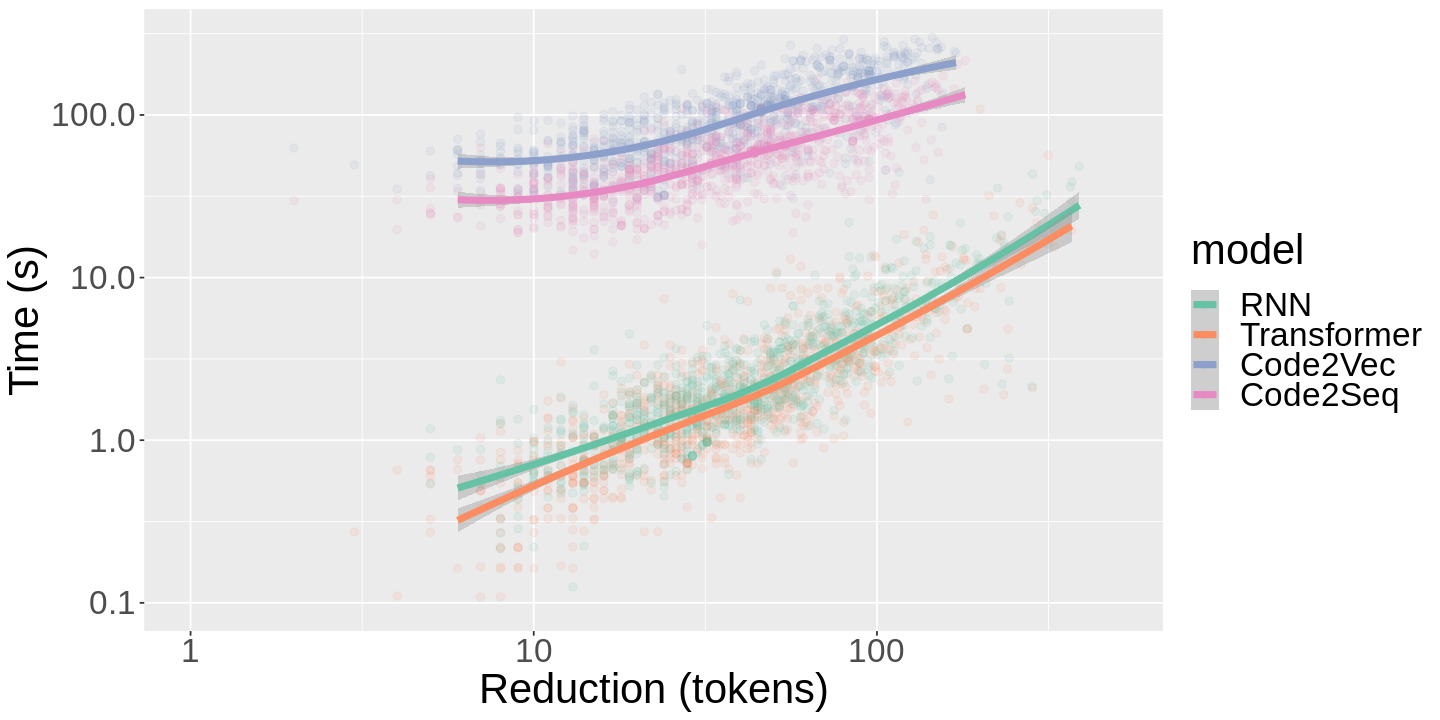}
    \caption{The reduction time (in seconds) vs. the number of tokens removed correlate roughly linearly, albeit with a substantial overhead on the \mnp task due to its parsing requirement. Note the log-scaling on both axes.}
    \label{fig:token-reduction-time}
\end{figure}

\subsection{RQ4: Cost of Reduction}
Column ``Time'' in Table~\ref{tbl:avg_summary} shows the average time spent on reduction of the input programs. 
The average time for reduction of the input programs in \vm task was lower than four seconds, while the average time needed for reducing an input program in the \mnp task was around $60$ and $107$ seconds in \cts and \ctv models, respectively. 
\Cref{fig:token-reduction-time} shows the detailed cost of running \approach, plotting the log-scaled reduction time in seconds versus the log-scaled total number of tokens reduced on the correctly predicted samples for all models.
In our experiments, the models for the \mnp task (\ctv and \cts) require a parseable program for making a prediction, while the models for the \vm task (\rnn and \tra) can work with any sequence of tokens.
Therefore, for the \mnp task, after each step of delta debugging, we first create the candidate program from tokens, then parse this candidate program to check whether it is valid, and finally preprocess it to make a prediction by a model. This adds significant overhead to \approach's runtime compared to the \vm task. 
Overall, most reductions in \mnp finished in less than four minutes, while the reductions in \vm generally completed in less than ten seconds.
This aside, all models scale roughly linearly in reduction time with the number of tokens that are eventually removed (which itself correlates strongly with initial size, \Cref{fig:percent-reduction}). 
The individual models within each task-category were quite similar in terms of throughput, with a minor effect observed based on each model's performance (e.g., RNNs are slower than Transformers).

The columns under ``\#DD Pass'' show that the search for minimal inputs in \mnp included creating and trying slightly more than $350$ intermediate results in the \DD, wherein, on average, around $37$ of them were syntactically correct and $33$ produced the same prediction as to the original input program.
Similarly, in models related to \vm, the reduction takes between $237$ and $241$ on average from which around $30$ intermediate reduced programs produce predictions similar to the original input program. 

\observation{Input program reduction based on tokens is efficient and can reach the minified input with the same prediction in a relatively small number of steps.}

\begin{figure}
\textbf{Original program:}
\begin{lstlisting}[style=custompython,emph={mtime},emphstyle=\color{deepred}]
def modified_time(self, name):
    try:
        file = open(self._path(mtime))
        mtime = float(file.readlines()[2])
        file.close()
    except:
        mtime = None
    return mtime
\end{lstlisting}

\textbf{RNN minimized example:}
\begin{lstlisting}[style=custompython,emph={mtime},emphstyle=\color{deepred}]
self, name)
        file(self._path(mtime)
        mtime(file()
        file
        mtime
    return mtime
\end{lstlisting}

\textbf{Transformer minimized example:}
\begin{lstlisting}[style=custompython,emph={mtime},emphstyle=\color{deepred}]
def modified_time(self, name)
        file(self(mtime
        mtime = file
        file
        mtime
    mtime
\end{lstlisting}
\caption{An example of two models learning different kinds of shortcuts. The misused variable is `mtime' (in red), incorrectly passed in place of `name' on line 3 of the original snippet.}
\label{lst:shortcut}
\end{figure}

\section{Discussion}
\label{sec:discussion}
The central motivation for using linguistics-inspired models on software has long been that source code is in a sense ``natural''; that is, it contains repetitive, predictable patterns much like natural text \cite{hindle2012naturalness}. 
Models that leverage this intuition have evolved from simple $n$-gram based models to complex neural networks, such as those used in this paper in recent years, and become applicable to (and successful at) a wide range of coding tasks.
Yet, they are still built and motivated with the same core premise of learning from programs in their entirety (punctuation and all) and in lexical order. Recent results have already begun to suggest that, in practice, these models may not be using much of their input programs for at least one task \cite{Wang:2021:demystifying, zheng2020probing}; however, this still focuses on otherwise natural, only simplified programs.

Our method makes no attempt at preserving any meaning or validity of the original program; we purely focus on the smallest amount of data with which our models could suffice. This allows us to show a new, stronger result: these four models across two tasks appear almost entirely indifferent to the naturalness of the provided code snippets -- remove as much as 90\% of tokens and both their predictions, and their confidence therein remain almost unchanged (\Cref{fig:reduction-prediction-score}). We discuss the reasons and implications of this observation using several examples in the remainder of this section.

\subsection{Explanations vs. Shortcuts}
\Cref{lst:shortcut} shows an example of a code snippet and its minified by an \rnn and a \tra model respectively. In this case, the (synthetically induced) error is fairly obvious; \lstinline{mtime} is passed to the \lstinline{_path} call where \lstinline{name} should be. If given the task to find such a bug, a programmer might notice this particular use-before-definition bug, which could then be \emph{explained} with a relatively small portion of the function, perhaps involving the unused parameter \lstinline{name} and \lstinline{mtime}'s use before its assignment.
Looking at the two minified programs, we see traces of these same explanations: the RNN model, which prioritizes lexically local patterns \cite{hellendoorn2020global}, primarily preserves the part of the program where \lstinline{mtime} is passed to \lstinline{_path} -- an unlikely combination given that  \lstinline{name} is in the immediate context. The Transformer, meanwhile, prefers to focus on the subsequent assignment to \lstinline{mtime}, which is out of place given its preceding use.

However, the existence of a short explanation does not naturally imply that much of the program is unnecessary, as it apparently does to our models. The two minified programs are arguably much harder to read than the original -- especially as our model would see them, without the original for reference, or the indentation that we added here -- yet our models seemed to have no more trouble predicting from them than the original. It is evident that our models have learned to take \emph{shortcuts}: they predominantly leverage simple syntactic patterns, quite literally to the exclusion of almost all else.
Our broader results show that this behavior is ubiquitous and extends across models and tasks (\cref{sec:results}).

These shortcuts still capture meaningful, natural programming patterns. Neither failing to use a parameter, nor the assignment after use,\footnote{The variable might be a field.} are necessarily syntactically invalid, but our models have clearly learned that they are sufficiently irregular to contemplate a bug. What is remarkable about our findings is not the absence of ``naturalness" as a whole, but the absence of \emph{macro-level naturalness}. Our models appear to have little to no care for the overall structure and content of the function, just for the presence or absence of specific patterns therein.

\subsection{Ramifications for Deep Learning}
Do our results then imply that deep learners are, in a sense, frauds, at least in software engineering applications? Not quite: it is well known that machine learners, neural networks especially, prefer to find simple explanations of their training data \cite{ilyas2019adversarial}, which often hinders their generalizability. The models in our case are doing nothing less: they are presented with a single, often obvious task and learn to solve it (arguably) as efficiently as possible. That does not invalidate the quality of their learning: in practice, there are myriad patterns to heed when predicting a vocabulary of hundreds of thousands of method names, or finding arbitrary variable misuse bugs in millions of functions. One interpretation of our results is that our models arrive at a set of simple ``explanations'' that encompasses nearly all these cases, such as how a variable should not be used before its assignment. This vocabulary is still large and diverse, so it remains a significant challenge for models to discover -- there are good reasons why model performance can differ vastly even on simple tasks \cite{hellendoorn2020global}.

We saw these differences in action, too: Transformer models are substantially better at leveraging ``global'' information from throughout the function than RNNs, which are largely (lexically) ``local''. Correspondingly, we saw that the latter permitted substantially more reduction of input programs, anecdotally because it mainly preserved features in the immediate context of the bug's location. Note that this is not a strength of the RNN: we are reducing programs to find out how much evidence these models used in their predictions. The ideal is neither the complete program nor virtually none of it.

Overall, the apparent indifference to macro-level naturalness (that functions as a whole are complete and well-formatted) is troubling. Much like prior findings, our demonstration that models rely on just a few features of their input naturally implies that they are highly vulnerable to perturbations in those inputs \cite{Vechev:AdversarialCode, rabin2020demystifying, Wang:2021:demystifying, zheng2020probing}. 
A natural question may be: what might prompt them to read code more holistically? One answer may come in the form of multi-task learning, in which a single model is trained on a wide variety of tasks \cite{kanade2019pre}.
In our analyses, we believe the models learned those salient patterns that helped achieve their singular objective; a mixture of diverse objectives might prevent such shortcuts altogether. Whether and how this works in practice is an open question. 
One risk may be that similar types of shortcuts are useful for many tasks, especially those based on  synthetically generated flaws -- a common practice in our field. If so, our approach should be able to elucidate this, and may well be able to serve more generally as an indicator of the complexity of a task and/or the degree of information used by a model, by using visualizations such as \Cref{fig:percent-reduction} that show the amount of input data required to accomplish a task on average.

\subsection{Accessible Prediction Attribution}
Prediction attribution in deep learners is a fast growing field, that attempts to relate the prediction of a neural network to a part of its input ~\cite{Attribution}.
Most current attribution approaches either require access to the trained model in full (``white-box'' methods), including its architecture and (hyper-)parameters, or try to approximate them, e.g., \cite{attribution-gradient-based,selvaraju2017grad:attribution:gradient}.
The methodology based on reduction that we used in this paper does not require any knowledge of either architecture or hyper-parameters of the models; it can be applied in a complete black-box fashion. This both makes it applicable to cases where internal information of models is unavailable, e.g. proprietary models, but also substantially eases its use. Deploying our model to a new pre-trained model only required knowing minimal constraints on the input (which tokens may not be reduced; whether the program must continue to be parseable), which are usually readily accessible. 

Moreover, current techniques in attribution and transparency usually require a certain level of knowledge about the characteristics of learning algorithms and reasoning about the model behaviors, which the average practitioner may not have sufficient time to acquire.
We envision that the application of \approach, if deployed directly to developers, would thus be more accessible to software engineers, especially those who have prior familiarity with the concept of test reduction for failing test cases in debugging, and this knowledge can be easily transferred to reducing \ciinputs in \approach.
Our results did show differences, sometimes significant ones, in reduction-behavior between various architectures (\eg \cref{lst:shortcut}), which may well be useful for experts to interpret. But, using \approach does not require such knowledge; the reduced input programs speak for themselves.
We plan to perform a user-study to evaluate the usability of our approach to the average practitioner.

\subsection{Limits of Extractive  Interpretation}
The \approach methodology proposed in this paper is best described as an \emph{extractive} interpretation approach. 
Extractive interpretation methods extract and present the important regions of the inputs, leaving it to the user to characterize the behavior of the model.
These approaches usually work well in cases where features are sparse and not correlated, so that highlighting one or more parts in the input provides enough insights about the model.
The high rate of reduction in most cases in our work may suggest that this approach is indeed applicable here and can provide sufficient insights about the behavior of the models -- although a user study is needed to validate that further.
In turn, the power of this approach would be limited on models that use complex or non-syntactic features such as the number of lines of code in programs, as the basis for their prediction. This would prevent the \ciinputs from being reduced significantly, or it might be difficult to make sense of the output and pinpoint the underlying important feature in the reduced programs.


\subsection{Impact of Smallest Atomic Unit Choice}
Choosing different smallest atomic units in the \DD algorithm can provide different, and potentially complementary, insights about the model.
For example, \ctv predicts the name of  the code snippet in Figure ~\ref{code:reduction} as `\lstinline{main}'. 
If we use hierarchical \DD wherein the smallest atomic unit of reduction is an AST node, the result would be ``\lstinline|void f (String args) {}|'' suggesting that the method signature and the argument name have had a large impact on the prediction. However, if we choose characters as the smallest atomic unit and employ \DD, the result is 
``\lstinline|d f(Sg[]r){em.s(C.D,"");Main(r);}|''
which provides a complementary view for the prediction that shows the importance of \texttt{Main} identifier used in the method body. Future work may extend our approach to new metrics and reduction strategies, which may well provide novel insights, especially in the future, more complex models that are more resistant to such simple reduction.

\begin{figure}
\begin{lstlisting}[style=custompython,emphstyle=\color{deepred},basicstyle=\footnotesize]
public static void f(String[] args) {
    System.setProperty(Constants.DUBBO_PROPERTIES_KEY,
        "conf/dubbo.properties"); 
    Main.main(args);
}
\end{lstlisting}
\caption{\ctv  correctly predicts the name of this method as `\texttt{main}'.}
\label{code:reduction}
\end{figure}

\section{Related Work}
\label{sec:related}

There has been a lot of work in the area of transparent or interpretable AI, computer vision, and natural language processing that focuses on understanding neural networks.
Interpretable, transparent machine learning has numerous benefits, including making predictions explainable (and thus more useful), using learned models to generate new insights, and improving the quality and robustness of the models themselves  \cite{samek2020interpretable}. 
These objectives are generally studied under the umbrella of ``explainable AI''. While some work studies the properties of neural models in general \cite{balduzzi2017shattered}, many studies are more ad-hoc, focusing on specific domains and tasks \cite{BigCodeSurvey}.

\subsection{Software Engineering}
There is a growing body of work in the domain of robust neural models for source code or code intelligence in general. 
Bui \etal~\cite{Nghi2019AutoFocus} evaluated the impact of a specific code fragment by deleting it from the original program, Rabin \etal~\cite{rabin2020demystifying} compared the performance of neural models and SVM models with handcrafted features and found comparable performance with a relatively small number of features.
Wang \etal~\cite{Wang:2021:demystifying} propose a mutate-reduce approach to find important features in the code summarization models.
Several other studies~\cite{rabin2019tnpa, rabin2021generalizability, Vechev:AdversarialCode, Reps:CodeRobustness} have evaluated the robustness of neural models of code under semantic-preserving program transformations and found that transformations can change the results of the prediction in many cases.
Finally, Sahil \etal~\cite{zheng2020probing}, published concurrently with this work, present a very similar approach to capturing vulnerability signals from a model’s prediction by applying prediction-preserving input minimization using delta-debugging. Their results are complementary to ours, further reinforcing the merit of the proposed method.

\subsection{Computer Vision and NLP}
The need for neural model interpretability and robustness was firmly established by Goodfellow \etal, who showed that a convolutional neural network could be tricked into changing its image classification into virtually any label by adding an imperceptible amount of targeted noise \cite{goodfellow2014explaining}. This produced a flood of research on both improving the robustness of neural networks and attacking current architectures \cite{carlini2017towards}, often in tandem. Importantly, comprehensive studies of robustness and adversarial perturbations have yielded rich scientific insights into the learning behavior of these models, both for vision and beyond \cite{ilyas2019adversarial}. 

The computer vision community has proposed many ways to visualize what parts of the input are most significant to a neural network, both to individual neurons and its output classification. A popular example, and similar to our approach, is occlusion analysis \cite{zeiler2014visualizing}, in which regions of interest are identified by evaluating the network's prediction when various parts of the input image are occluded. Locations at which prediction accuracy rapidly drops when removed are likely especially integral to the prediction. For linguistic expressions, input perturbations are usually less obvious: while certain words (such as stop words) may safely be removed without altering the meaning of a sentence, more general changes quickly risk producing very different inputs. Recent input-related methods rely on synonym datasets and swap out similar names to ensure that they generate semantically similar phrases \cite{ren2019generating,jin2019bert}.
Our work shows that, at least for current models and tasks, this is significantly less of a concern in software engineering, where many tokens can be removed with little consequence.

\section{Threats to Validity}
\label{sec:validity}
We evaluated \approach on four neural models and two tasks, trained on millions but tested on a few thousands of random samples from their datasets. As such, our findings may naturally not extend to other inputs, models, and domains. Nevertheless, we argue both that our analysis is broad, spanning more models and domains than comparable work \cite{Wang:2021:demystifying}, and that the design of our approach is applicable to many other problem settings in our field, which commonly take tokens and ASTs as inputs to yield a single, or few outputs, all covered by the models in this work.
To ensure software quality, we used the tools and datasets shared by the original developers of the models, each from public repositories used by dozens of developers and cited in multiple studies. For our input reduction, we adapted Zeller \etal's implementation of \DD \cite{Zeller:2002:simplifying}, which has been widely used in the industry and other research studies over decades.
\section{Conclusion}
\label{sec:conclusion}
We proposed \approach, a simple, model-agnostic methodology for interpreting a wide range of code intelligence models, which works by reducing the size of \ciinputs using the well-known delta-debugging algorithm. We apply \approach to four popular neural code intelligence models across two datasets and two tasks, showing that our method can significantly reduce the size of \ciinputs while preserving the prediction of the model, thereby exposing the most significant input features to the various models.
Our results hint at the idea that these models often use just a few simple syntactic shortcuts in their prediction.
This sets the stage for broader use of transparency-enhancing techniques to better understand and develop neural code intelligence models.

\balance
\bibliography{bib,amin}
\bibliographystyle{acm}

\end{document}